\documentclass[sigconf]{acmart}
\usepackage{multirow}
\AtBeginDocument{%
  }

\setcopyright{acmlicensed}
\copyrightyear{2024}
\acmYear{2025}
\acmDOI{XXXXXXXXX}
\acmConference[WWW'25 Companion]{}{Apr 28 - May 02, 2025}{Sydney, Australia}
\acmISBN{978-1-4503-XXXX-X/18/06}

\begin{document}

\title{GraphSeqLM: A Unified Graph Language Framework for Omic Graph Learning}

\author{Heming Zhang}
\email{hemingzhang@wustl.edu}
\affiliation{%
  \institution{Washington University in St. Louis}
  \city{St. Louis}
  \state{MO}
  \country{USA}
}

\author{Di Huang}
\email{di.huang@wustl.edu}
\affiliation{%
  \institution{Washington University in St. Louis}
  \city{St. Louis}
  \state{MO}
  \country{USA}
}

\author{Yixin Chen}
\email{ychen25@wustl.edu}
\affiliation{%
  \institution{Washington University in St. Louis}
  \city{St. Louis}
  \state{MO}
  \country{USA}
}

\author{Fuhai Li}
\email{fuhai.li@wustl.edu}
\authornote{Correpsonding author}
\affiliation{%
 \institution{Washington University in St. Louis}
 \city{St. Louis}
 \state{MO}
 \country{USA}
}

\renewcommand{\shortauthors}{Zhang et al.}

\begin{abstract}
The integration of multi-omic data is pivotal for understanding complex diseases, but its high dimensionality and noise present significant challenges. Graph Neural Networks (GNNs) offer a robust framework for analyzing large-scale signaling pathways and protein-protein interaction networks, yet they face limitations in expressivity when capturing intricate biological relationships. To address this, we propose Graph Sequence Language Model (GraphSeqLM), a framework that enhances GNNs with biological sequence embeddings generated by Large Language Models (LLMs). These embeddings encode structural and biological properties of DNA, RNA, and proteins, augmenting GNNs with enriched features for analyzing sample-specific multi-omic data. By integrating topological, sequence-derived, and biological information, GraphSeqLM demonstrates superior predictive accuracy and outperforms existing methods, paving the way for more effective multi-omic data integration in precision medicine \footnote{https://github.com/FuhaiLiAiLab/GraphSeqLM}.
\end{abstract}



\begin{teaserfigure}
\includegraphics[width=\textwidth]{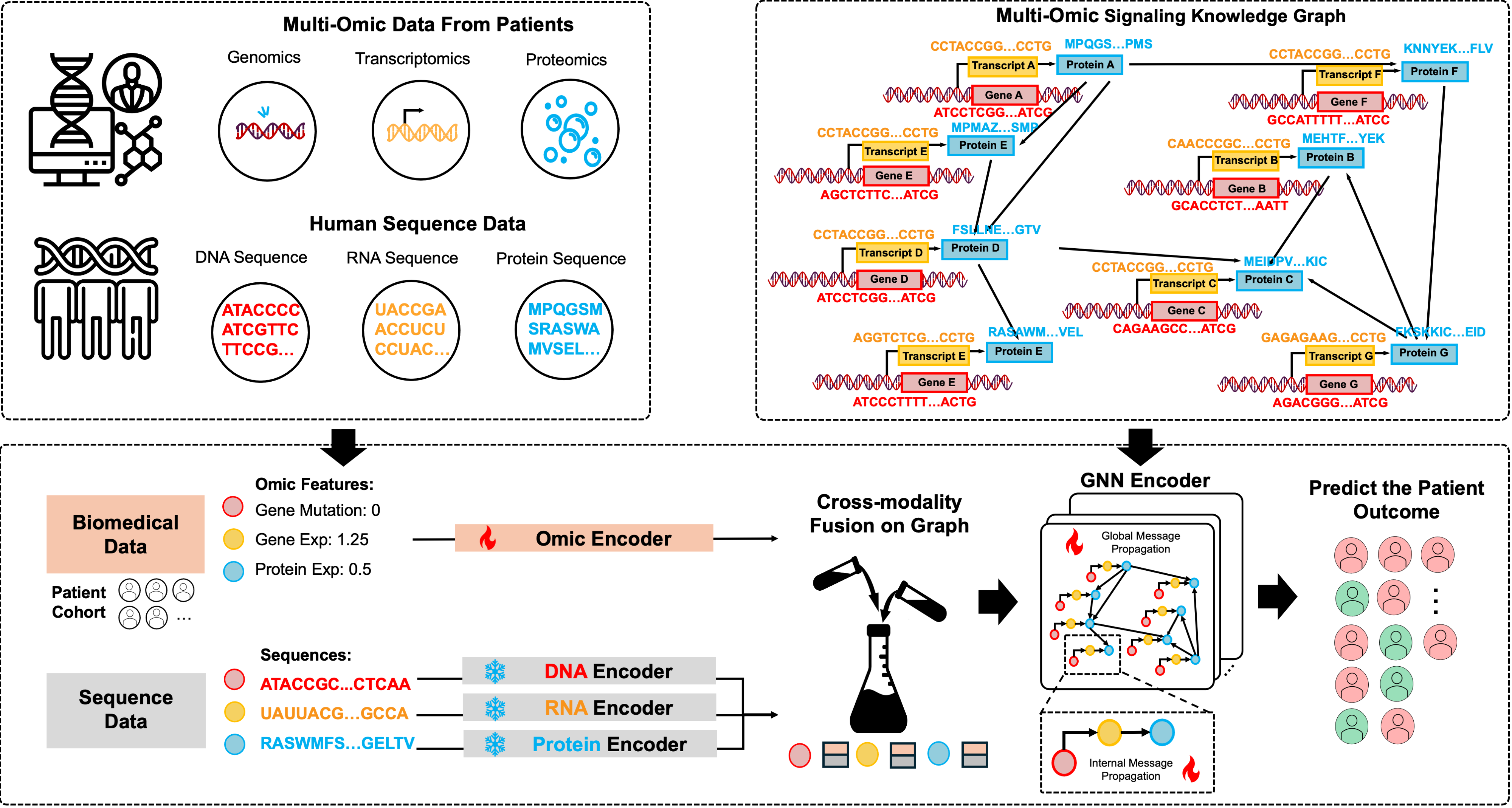}
\caption{Overview of GraphSeqLM: The framework integrates multi-omic data and biological sequences into a Multi-Omic Signaling Knowledge Graph, utilizing unified graph language encoder to predict patient outcomes}
\Description{}
\label{fig:teaser}
\end{teaserfigure}


\maketitle
\section{Introduction}
Graph neural networks (GNNs) have emerged as powerful tools for analyzing graph-structured and relational data, which are prevalent in domains like life science and social networks. In life science and precision medicine, the advent of high throughput multi-omic data, characterizing the variations in genomic, epigenomic, transcriptomic, and proteomic, has enabled deeper and multi-view insights into the pathogenesis of complex diseases. Integrating complex, high-dimensional, and noisy multi-omic datasets remains a significant challenge. Graph-based models, such as Graph Neural Networks (GNNs), have demonstrated effectiveness in addressing this challenge by providing an integrative and interpretable framework for multi-omic data analysis. These models inherently represent complex disease signaling systems through large-scale signaling pathways or protein-protein interaction networks, capturing intricate interactions among extensive sets of genes and proteins, whose activation states are defined by multi-omic features \cite{mogonet2021, Zhao2022MODIGIM, Zhang2024Systems, Dong2024, Yang2023MRGCNCS, Li2022MoGCNAM}. Advances such as adaptive graph learning \cite{Dong2023IntegrationOM} and graph attention networks (GATs) for multi-dimensional data integration \cite{Zhao2022MODIGIM, velivckovic2017graph} have further enhanced the capabilities of these approaches. Additionally, the development of multi-omic signaling (MOS) knowledge graphs (KGs) has enabled more integrative and interpretable modeling of biological systems \cite{Zhang2024MosGGen}. However, graph encoders still encounter limitations in expressivity when learning and mining patterns within large, complex signaling graphs that involve dense interactions among numerous proteins and genes. A promising strategy to address these challenges is feature augmentation through the incorporation of pre-computed features \cite{abboud2020surprising, lim2022sign}.
Simultaneously, Large Language Models (LLMs) have revolutionized natural language processing (NLP) with their remarkable versatility, scalability, and groundbreaking potential across a wide range of applications. Beyond their textual successes, LLMs have demonstrated extraordinary capabilities in representing complex sequences, including biological sequences such as DNA, RNA, and protein sequences. In this study, we introduce a novel Graph Language Framework, GraphSeqLM, which integrates LLM-generated biological sequence embeddings with Graph Neural Networks (GNNs). This approach encodes the inherent biological and structural properties of DNA, RNA, and protein sequences, serving as a feature enhancement mechanism. By incorporating these embeddings with sample-specific multi-omic data, GraphSeqLM integrates topological, biological, and sequence-derived information, enabling the model to capture intricate biological patterns with heightened predictive accuracy. This innovative fusion of LLM-driven sequence embeddings and multi-omic data within the framework of Graph Neural Networks (GNNs) represents a novel advancement in the field, surpassing the performance of existing graph-based approaches. Comprehensive evaluation results underscore the efficacy and superiority of our proposed methodology.

\section{Methodology}
\subsection{Problem Formulation}
This section outlines a detailed methodology addressing the integration of biological sequence data, including DNA, RNA, and protein sequences, into multi-omic datasets within the graph language framework. The proposed integration aims to enhance the representation  of complex biological networks by leveraging sequence embeddings derived from LLMs. By combining sequence-level information with multi-omic profiles such as genomic, transcriptomic, and proteomic, this approach seeks to improve the predictive capabilities of graph-based models. 

Specifically, given the bulk-seq multi-omic dataset of $\mathcal{X}=\{\mathcal{X}^{(g)}$, $\mathcal{X}^{(t)}, \mathcal{X}^{(p)}\}$, where $\mathcal{X}^{(g)}$ is the genomic feature set, $\mathcal{X}^{(t)}$ is the transcriptomic feature set and $\mathcal{X}^{(p)}$ is the proteomic feature set, these multi-omic features were integrated into MOS-KG by existed gene regulatory network, Kyoto Encyclopedia Genomes and Genes (KEGG\cite{kanehisa2000kegg}), resulting with the sequence-attributed knowledge graph $\mathcal{G}=(\mathcal{V, E})$. In this knowledge graph, $\mathcal{V} = \{\mathcal{V}^{(g)}, \mathcal{V}^{(t)}, \mathcal{V}^{(p)}\}$ represents the set of vertices, where $|\mathcal{V}^{(g)}| = n^{(g)}$, $|\mathcal{V}^{(t)}| = n^{(t)}$, and $|\mathcal{V}^{(p)}| = n^{(p)}$. This graph can be decomposed into two subgraphs: $\mathcal{G}^{(\text{in})} = (\mathcal{V}^{(\text{in})}, \mathcal{E}^{(\text{in})})$ and $\mathcal{G}^{(\text{PPI})} = (\mathcal{V}^{(\text{PPI})}, \mathcal{E}^{(\text{PPI})})$. Here, $\mathcal{G}^{(\text{in})}$ captures the internal signaling processes for protein translation, with $\mathcal{V} = \mathcal{V}^{(\text{in})}$ and $|\mathcal{V}^{(\text{in})}| = n = n^{(g)} + n^{(t)} + n^{(p)}$, while $\mathcal{G}^{(\text{PPI})}$ represents the gene regulatory network, structured around protein-protein interactions (PPI), where $\mathcal{V}^{(\text{PPI})} = \mathcal{V}^{(p)}$. Additionally, the patient multi-omic feature set is denoted as $\mathcal{X} = \{X^{(1)}, X^{(2)}, \ldots, X^{(m)}, \ldots, X^{(M)}\}$, with $X^{(m)} \in \mathbb{R}^n$, and the patient label matrix is represented as $Y \in \mathbb{R}^M$. 

Aside from the multi-omic feature, the sequence dataset, $\mathcal{S}=\{S^{(g)}, S^{(t)}, S^{(p)}\}$, where $S^{(g)}=[s^{(1)}, s^{(2)}, \cdots, s^{(n^{(g)})}]$, $S^{(g)}\in{\mathbb{R}^{n^{(g)}}}$, is the DNA sequences for gene nodes $\mathcal{V}^{(g)}$, $S^{(t)}=[s^{(1)}, s^{(2)},$ $\cdots, s^{(n^{(t)})}]$ $S^{(t)}\in{\mathbb{R}^{n^{(t)}}}$, is the RNA sequences for transcript nodes $\mathcal{V}^{(t)}$ and $S^{(p)}=[s^{(1)}, s^{(2)},$ $ \cdots,$ $s^{(n^{(p)})}]$, $S^{(p)}\in{\mathbb{R}^{n^{(p)}}}$, is the protein sequences for protein nodes $\mathcal{V}^{(p)}$, was included to improve the graph expressivity for generating the sequence-attributed multi-omic signaling knowledge graph. The ultimate goal is to construct a graph language model, $f(\cdot)$, that leverages the integrated features $\mathcal{X}$, sequences $\mathcal{S}$, and graph edges $\mathcal{E}$ to predict patient outcomes. The model is defined as follows:

\begin{equation}
\hat{Y}=f(\mathcal{X}, \mathcal{S}, \mathcal{E})
\end{equation}
, where $\hat{Y} \in \mathbb{R}^{M}$ denotes the predicted patient outcomes.

\subsection{Infrastructure of GraphSeqLM}
In details, the proposed model, GraphSeqLM, consists of three core components. The Sequence Language Encoders extract embeddings from biological sequences such as DNA, RNA, and proteins, capturing key sequence-level features. These embeddings are fused with multi-omic data within the constructed multi-omic Signaling Knowledge Graph (MOS-KG) to integrate diverse modalities. Finally, a GNN Encoder leverages this enriched graph representation to predict patient outcomes effectively.

\textbf{Sequence Language Encoders} To embed the sequence data for each node in the generated MOS-KG, $\mathcal{G}$, different sequence language models were considered for various node types, i.e., gene sequences, transcript sequences and protein sequences leverage DNA, RNA and protein encoders. Therefore, sequence embeddings will be generated by

\begin{equation}
Q^{(g)} = \text{DNASequenceEncoder}(S^{(g)})
\end{equation}
\begin{equation}
Q^{(t)} = \text{RNASequenceEncoder}(S^{(t)})
\end{equation}
\begin{equation}
Q^{(p)} = \text{ProteinSequenceEncoder}(S^{(p)})
\end{equation}
, where $Q^{(g)}\in{\mathbb{R}^{n^{(g)}\times{d^{(l)}}}}$, $Q^{(t)}\in{\mathbb{R}^{n^{(t)}\times{d^{(l)}}}}$, $Q^{(p)}\in{\mathbb{R}^{n^{(p)}\times{d^{(l)}}}}$ are gene, transcript and protein sequence embeddings. Thus, node sequence embeddings $Q=[Q^{(g)}, Q^{(t)}, Q^{(p)}]$ ($Q\in{\mathbb{R}^{n\times{d^{(l)}}}}$) were generated.

\textbf{Cross-modality Fusion on Graph} Given the sequence embeddings, the nodes on MOS-KG will be augmented with the initial feature depending on the node types by 

\begin{equation}
H^{(m)} = \text{OmicEncoder}(X^{(m)})
\end{equation}
\begin{equation}
H'^{(m)} = \text{CrossModalityEncoder}(Q, H^{(m)})
\end{equation}

, where $X^{(m)}=[X^{(m)}_{t}, X^{(m)}_{g}, X^{(m)}_{p}]$ ($X^{(m)}\in{\mathbb{R}^{n}}$) is the multi-omic feature matrix and $X^{(m)}_{g}\in{\mathbb{R}^{n^{(g)}}}$, $X^{(m)}_{t}\in{\mathbb{R}^{n^{(t)}}}$, $X^{(m)}_{p}\in{\mathbb{R}^{n^{(p)}}}$ are genomic, transcriptomic and proteomic feature matrices, $H^{(m)}\in{\mathbb{R}^{n\times{d^{(l)}}}}$ is the multi-omic embeddings and $H'^{(m)}\in{\mathbb{R}^{n\times{d^{(l')}}}}$ is the merged embeddings on MOS-KG. Therefore, the genomic features and gene sequences, transcriptomic features and transcript sequences, and proteomic features and protein sequences will be merged on the gene nodes, $\mathcal{V}^{(g)}$, transcript nodes, $\mathcal{V}^{(t)}$ and protein nodes, $\mathcal{V}^{(p)}$, respectively.

\textbf{Graph Neural Network Encoder} With the sequence augmented multi-omic node embeddings, $\mathcal{H}=\{H'^{(1)}, H'^{(2)}, \cdots,$ $H'^{(m)},$ $\cdots, H'^{(M)}\}$,  the information will be propagated on the internal path to make sure the information diffused on the protein nodes with consideration of the biological meaning of protein translation process by

\begin{equation}
Z^{(m)}=\text{GNN}_{\text{in}}(H'^{(m)}, \mathcal{E}^{\text{(in)}})
\end{equation}

, where $\text{GNN}_{\text{in}}$ is the graph-based internal message propagation to ensure the signaling flow for protein translation and $Z^{(m)}\in{\mathbb{R}^{n\times{d}}}$ denotes the node internal embeddings. Afterwards, the global message propagation will be made by

\begin{equation}
Z'^{(m)}=\text{GNN}_{\text{global}}(Z^{(m)}, \mathcal{E})
\end{equation}

, where $\text{GNN}_{\text{global}}$ is also the graph-based global message propagation for signaling flow over whole regulatory network and $Z'^{(m)}\in{\mathbb{R}^{n\times{d}}}$ denotes the node global embeddings.

\textbf{Patient Outcomes Prediction} Finally, with the global node embeddings, $\mathcal{Z}=\{Z'^{(1)}, Z'^{(2)},\cdots,Z'^{(m)},\cdots, Z'^{(M)}\}$, the graph prediction will be made by

\begin{equation}
O^{(m)}=\text{MLP}(\text{AVG}(Z'^{(m)}))
\end{equation}
\begin{equation}
\hat{Y}^{(m)}=\text{arg max}(O^{(m)})
\end{equation}

, where $\text{AVG}$ is the average aggregation function, MLP denotes the multi-layer perceptron for linear transformation to predict patient outcomes for this classification task and $O^{(m)}\in{\mathbb{R}^{c}}$. The final predicted outcome will be extracted by maximum function with $\hat{Y}^{(m)}\in{\{0,1,\cdots,c-1\}}$.

\section{Experiment and Evaluation}
\subsection{Dataset Collection and Baseline Models}
The multi-omic data for cancer patients were obtained from the UCSC Xena browser \cite{goldman2018ucsc}. After integrating the multi-omic datasets and excluding cancer samples with imbalanced or insufficient data, a final cohort of 826 samples across 6 cancer types was curated (see \textbf{Table~\ref{tab:cancer_types_details}}), each labeled with overall survival status. This dataset forms the basis for a binary classification task ($c=2$). Additionally, biological sequence data were retrieved from Ensembl \cite{harrison2024ensembl} using the Python API, and the corresponding biological regulatory networks were sourced from KEGG \cite{kanehisa2000kegg}.

\begin{table}[h!]
    \centering
    \caption{Cancer types and sample details}
    \label{tab:cancer_types_details}
    \small % Use smaller font size
    \renewcommand{\arraystretch}{0.8} % Reduce row spacing
    \begin{tabular}{c c c c}
        \toprule
        Cancer Types & Total Samples & \# of Survival & \# of Death \\
        \midrule
        BLCA & 116 & 60 & 56 \\
        GBM & 67 & 49 & 18 \\
        LUAD & 172 & 79 & 93 \\
        LUSC & 103 & 47 & 56 \\
        SKCM & 59 & 17 & 42 \\
        STAD & 309 & 121 & 188 \\
        \midrule
        Overall & 826 & 373 & 453 \\
        \bottomrule
    \end{tabular}
\end{table}

To evaluate improvements over existing GNN models, baseline architectures including GCN \cite{kipf2016semi}, GAT \cite{velivckovic2017graph}, GIN \cite{xu2018powerful}, and UniMP \cite{shi2020masked} were used. For GraphSeqLM, DNAGPT \cite{zhang2023dnagpt} and ProtGPT2 \cite{ferruz2022protgpt2} served as GPT-based encoders for DNA, RNA, and protein sequences, where DNA and RNA shared the same encoder by substituting thymine (T) with uracil (U) for RNA.

\subsection{Experiment Settings and Evaludation}
The 5-fold cross-validation approach was employed to partition each multi-omic dataset for each cancer type and two evaluation metrics, prediction accuracy and F1-score were utilized. As shown in \textbf{Table~\ref{tab:model_performances}}, the proposed model, GraphSeqLM, consistently outperforms other graph-based encoder models, demonstrating its superior effectiveness. These results emphasize the strength of GraphSeqLM in integrating sequence embeddings and multi-omic data, improving predictive capabilities in complex biological systems.

\begin{table*}[h!]
    \centering
    \caption{Model performance (accuracy and F1-scores) for different cancer datasets}
    \label{tab:model_performances}
    \small % Reduce font size
    \renewcommand{\arraystretch}{0.9} % Reduce row spacing
    \begin{tabular}{c c c c c c c}
        \toprule
        \textbf{Cancer Types} & \textbf{Metrics} & \textbf{GAT} & \textbf{GCN} & \textbf{GIN} & \textbf{UniMP} & \textbf{GraphSeqLM} \\
        \midrule
        \multirow{2}{*}{BLCA} & Accuracy & $0.7761 \pm 0.0552$ & $0.7152 \pm 0.0271$ & $0.7156 \pm 0.0379$ & $0.7583 \pm 0.0737$ & $\mathbf{0.7841 \pm 0.0628}$ \\
                              & F1 Score & $0.7138 \pm 0.263$ & $0.6669 \pm 0.2368$ & $0.617 \pm 0.3497$ & $0.7084 \pm 0.1754$ & $\mathbf{0.7743 \pm 0.106}$ \\
        \midrule
        \multirow{2}{*}{GBM} & Accuracy & $0.8319 \pm 0.1274$ & $0.7879 \pm 0.1261$ & $0.833 \pm 0.1146$ & $0.7868 \pm 0.1492$ & $\mathbf{0.8945 \pm 0.0671}$ \\
                              & F1 Score & $0.8884 \pm 0.0884$ & $0.8634 \pm 0.0867$ & $0.8839 \pm 0.0861$ & $0.8591 \pm 0.0991$ & $\mathbf{0.9145 \pm 0.0618}$ \\
        \midrule
        \multirow{2}{*}{LUAD} & Accuracy & $\mathbf{0.6983 \pm 0.0557}$ & $0.6803 \pm 0.0452$ & $0.6867 \pm 0.0714$ & $0.6689 \pm 0.0706$ & $0.6746 \pm 0.0352$ \\
                              & F1 Score & $0.657 \pm 0.058$ & $0.6618 \pm 0.0727$ & $\mathbf{0.677 \pm 0.0335}$ & $0.6162 \pm 0.1238$ & $0.5918 \pm 0.1688$ \\
        \midrule
        \multirow{2}{*}{LUSC} & Accuracy & $0.7186 \pm 0.0385$ & $0.6995 \pm 0.0592$ & ${0.7281 \pm 0.0264}$ & $0.7086 \pm 0.0078$ & $\mathbf{0.7376 \pm 0.0287}$ \\
                              & F1 Score & $0.5889 \pm 0.2214$ & $0.4734 \pm 0.348$ & ${0.6234 \pm 0.2321}$ & $0.4921 \pm 0.2727$ & $\mathbf{0.6245 \pm 0.2203}$ \\
        \midrule
        \multirow{2}{*}{SKCM} & Accuracy & $0.7788 \pm 0.0975$ & $0.7636 \pm 0.1065$ & $0.8303 \pm 0.0593$ & $0.8121 \pm 0.0756$ & $\mathbf{0.8621 \pm 0.0836}$ \\
                              & F1 Score & $0.2943 \pm 0.2755$ & $0.2 \pm 0.2981$ & $0.4633 \pm 0.2931$ & $0.4433 \pm 0.2934$ & $\mathbf{0.5332 \pm 0.3706}$ \\
        \midrule
        \multirow{2}{*}{STAD} & Accuracy & $0.6701 \pm 0.0734$ & $0.6605 \pm 0.0835$ & $0.6604 \pm 0.0644$ & $\mathbf{0.6831 \pm 0.0619}$ & $0.6734 \pm 0.0766$ \\
                              & F1 Score & $0.3808 \pm 0.2027$ & $0.2767 \pm 0.2377$ & $\mathbf{0.4132 \pm 0.2433}$ & $0.4034 \pm 0.1576$ & $0.3319 \pm 0.1847$ \\
        \bottomrule
    \end{tabular}
\end{table*}

\section{Discussions}
The integration of biological sequence data into multi-omic datasets using the proposed GraphSeqLM framework highlights significant advancements in modeling complex biological networks. By leveraging embeddings from LLMs, GraphSeqLM captures sequence-level properties of DNA, RNA, and proteins, which complement multi-omic features such as genomic, transcriptomic, and proteomic data. This cross-modality fusion enhances the expressive power of GNNs, enabling them to better represent intricate interactions in large-scale signaling pathways and protein-protein interaction networks.

Compared to traditional graph-based approaches, GraphSeqLM effectively overcomes limitations in graph encoder expressivity by incorporating enriched node features through sequence embeddings. This integration allows the model to improve predictive accuracy for outcome prediction. Comprehensive evaluation results demonstrate that GraphSeqLM outperforms existing GNN-based frameworks on most datasets, underscoring its potential for advancing precision medicine and facilitating deeper insights into disease mechanisms through multi-omic data analysis.

\begin{acks}
This research was partially supported by NLM 1R01LM013902-01A1.
\end{acks}

\bibliographystyle{ACM-Reference-Format}
\bibliography{sample-base}

\end{document}